# The quest for LEDBAT fairness

Giovanna Carofiglio*, Luca Muscariello†, Dario Rossi‡ and Silvio Valenti‡
* Bell Labs, Alcatel-Lucent, France, giovanna.carofiglio@alcatel-lucent.com
† Orange Labs, France, luca.muscariello@orange-ftgroup.com
‡ Telecom ParisTech, France, firstname.lastname@enst.fr

*Abstract*—BitTorrent developers have recently introduced a new application layer congestion control algorithm based on UDP framing at transport layer and currently under definition at the IETF LEDBAT Working Group. LEDBAT is a delay-based protocol which aims at offering a "lower than Best Effort" data transfer service, with a lower priority with respect to elastic TCP and interactive traffic (e.g., VoIP, game). However, in its current specification, LEDBAT is affected by a late-comer advantage: indeed the last flow arriving at the bottleneck is more aggressive due to a wrong estimation of the base delay and finally takes over all resources. In this work, we study several solutions to the late-comer problem by means of packet level simulations and simple analysis: in the investigation process, we individuate the root cause for LEDBAT unfairness and propose effective countermeasures.

## I. INTRODUCTION

Congestion control algorithms for the transfer of data on the Internet have long been studied: thus, the issue of congestion control is definitively not a new topic. However, the fact that BitTorrent has recently replaced TCP by a new algorithm for data transfers renews the relevance of the subject – as BitTorrent is among the most popular P2P applications and generates a significant amount of Internet traffic.

The initial misunderstanding of the protocol objectives (the announce caused an unmotivated buzz about the imminent Internet meltdown [1], soon officially denied [2]), pushed BitTorrent to co-chair a IETF Working Group for the development of the new protocol, named "Low Extra Delay Background Transfer" (LEDBAT). The LEDBAT protocol [3] is designed to be effective for P2P file-sharing, efficiently exploiting available bandwidth but at the same time avoiding self-congestion at the access. LEDBAT goal is to provide a "lower than Best Effort" data-transfer service, yielding to elastic TCP and interactive traffic like VoIP or gaming. To this purpose, LEDBAT implements a combined delay and loss based congestion control: the delay-based component is inspired by Vegas [4], while reaction to losses is the same of traditional TCP versions (i.e. multiplicative drop of the congestion window). In absence of packet losses, and with the goal of preventing them, LEDBAT basically monitors delay variation on the forward path. It adjusts its congestion window trying to keep the queuing delay as close as possible to a predefined *target* in order to guarantee an efficient utilization of the resource.

Yet this implies that, if two flows have a different measure of the base delay, they may estimate a different queuing delay: in particular, the flow with a lower base delay will sense a larger queuing delay, with consequent unfairness in congestion window update. More precisely, a *late-comer advantage* arises: while the first flow arriving at an empty bottleneck correctly measures the delay, a subsequent flow accounts the queuing delay of the first one in its base delay measurement, thus setting a higher target delay. Therefore, the second flow will aggressively take over the target share of the first one, eventually entering a possibly persistent unfair state.

In this work, we investigate this problem by means of packet level simulation, and outline several alternative solutions, tailored for the LEDBAT protocol. The first group of solutions only tries to *ameliorate the base delay measurement*. First, as suggested in the LEDBAT WG [5], we implement *random pacing* of packets belonging to the same window: this should allow flows to gather different delay samples and possibly converge to a similar view of the base delay. Second, we propose to use *TCP's slow-start* at the very beginning of LEDBAT flows: by filling the buffer, slow-start likely induces losses on already present flows, which drain the queue empty and leave a chance for new-comers to gather a correct measure of the base delay. The second group of solutions instead *addresses the window decrease decisions*, which represent a more fundamental issue. As third solution, we thus suggest introducing (infrequent) *random drops* of LEDBAT sender window, as a means to break unfair states and to de-correlate flow decisions. Fourth, we propose to replace the LEDBAT additive decrease with a *multiplicative decrease*: we indeed expect the abrupt reduction of the throughput of flows to empty the buffer and again allow late-comers to measure the real base-delay.

To summarize our main results, first we find that random pacing of packets within a window is ineffective in solving the fairness issue. Slow-start, instead, represents only a partial solution: fairness among LEDBAT flows improves, but this technique also causes problems to the performance of other kinds of traffic (e.g., TCP, VoIP, game), which LEDBAT strives against by design. On the other hand, the introduction of random drops of the LEDBAT sender window and the use of multiplicative decrease are both valid solutions. Interestingly, a performance tradeoff exists between them: the former maximizes efficiency metrics, while the latter maximizes the fairness performance.

Although both solutions have their merits, the latter may be more appropriate in this context. Indeed, we notice on the one hand that LEDBAT targets a lower than best-effort solution, and can thus tolerate a slight efficiency loss. On the other hand, we also point out that the fairness property may assist decisions taken at the P2P overlay level, such as peer selection, which are crucial to the overall system performance.

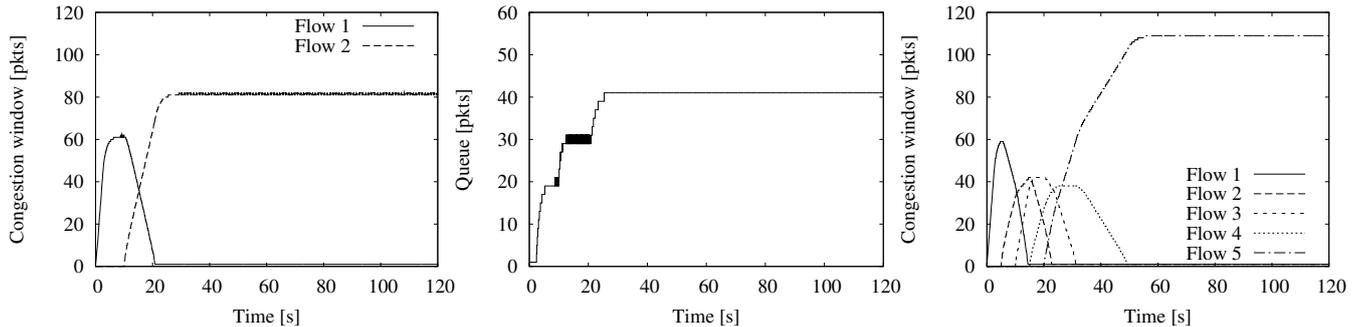

Fig. 1. Late-comer advantage: two flow scenario with queue evolution and multiple flow scenarios

## II. RELATED WORK

Congestion control studies date back to the 80's, therefore a thorough overview of the literature on this topic is out of scope of the present paper. Here, we simply recall that congestion control protocols can be divided into two categories according to the congestion indicator to which the protocol reacts. *Loss-based* protocols decrease their window when packet losses are detected, while *delay-based* protocols modulate the congestion window according to the queuing delay measured.

A number of loss and delay-based protocols exist, e.g. TCP Compound [6] or TCP Illinois [7], whose objectives are however different than LEDBAT's: in fact these protocols target higher efficiency rather than lower priority. Lower than TCP priority is instead the goal of TCP-LP [8], TCP-NICE [9] and 4CP [10], with whom LEDBAT shares some design aspects: however, the comparison of these protocols has not been explored to date, and remains an interesting open point.

BitTorrent studies have only recently [11], [12] started digging the LEDBAT issue, as early work focused on other aspects, such as torrent popularity [13], mechanisms for proximity aware peer selection [14], robustness of tit-for-tat mechanism [15] and models for the completion time of a swarm [16]. In our previous work [11], we investigate the LEDBAT congestion control by means of experimental measurement in a controlled test-bed, whereas we conducted a preliminary analysis of LEDBAT performance by means of simulation in [12]. As opposite to [11], [12], that exploit different approaches to evaluate the performance of the LEDBAT protocol "as is", in this work we instead focus on a specific weakness of the LEDBAT algorithm, the late-comer issue, and propose modifications apt at solving it.

## III. LEDBAT OVERVIEW AND FAIRNESS ISSUE

In this section, we briefly overview how the LEDBAT congestion control algorithm works and illustrate the conditions under which the fairness issue arises. For lack of space, we focus on protocol aspects relevant to the fairness problem, and refer the reader to [3] for a detailed description of the protocol.

As earlier mentioned, LEDBAT is a delay and loss-based protocol, designed to provide a low priority transport with respect to TCP. It decreases its rate when queuing delay grows up, before congestion arises and packets are eventually lost. Therefore, it mainly operates as a delay-based protocol where the congestion window increase/decrease is driven by the estimated queuing delay. In order to gather a measure of the delay on the communication path, each packet is time-stamped by both the source TX and the destination RX. TX maintains also a minimum over all delay measurements, the *base* delay, which represents an estimation of the propagation delay. Any further delay is then considered as queuing delay. Notice that, though TX and RX are not synchronized, since LEDBAT just considers *variations* of the delay, measurement errors due to clock offset and skew are canceled out by the difference.

The evolution of the congestion window, $W(t)$, is driven in LEDBAT by a linear controller, with a slope that depends on the difference between the target $\tau$ and the estimated queuing delay $q(t)$. The controller goal is to introduce a small non-zero *target* queuing delay, $\tau$, at the bottleneck. In the following, we select the gain parameter present in the draft as equal to $1/\tau$, in order to have at most the same increase slope of standard TCP Reno. Thus, at each packet arrival, $W(t)$ is adjusted as follows:

$$W(t+1) = \begin{cases} \frac{1}{2}W(t) & \text{if packet loss,} \\ W(t) + \frac{1}{W(t)}\frac{\tau - q(t)}{\tau} & \text{otherwise} \end{cases} \quad (1)$$

The queuing delay, $q(t)$ is measured as the difference between the current delay estimation and the minimum delay observed, i.e. the *base* delay. From (1) it can be seen that once the target is reached, the LEDBAT sender persists in this state unless other traffic (or a packet loss) perturbs the delay measurement.

To illustrate the late-comer unfairness, let us consider Fig. 1, which depicts an unfair situation for two (a)-(b) or many (c) LEDBAT flows sharing the same bottleneck. In Fig. 1-(a) we show the congestion window evolution of two flows arriving at the bottleneck respectively at time $t_1 = 0$ and $t_2 = 10\,\text{s}$. Fig. 1-(b) reports the queue size for this simple scenario. At the beginning the first flow correctly measures the base delay, sets its target delay to $\tau$ and starts increasing its sending rate until it finally contributes to queuing. The increasing phase stops at about $t = 5$ s, when the amount of packets in the queue is equal to 20 packets, which corresponds to a delay equal to the target $\tau = 25$ ms.

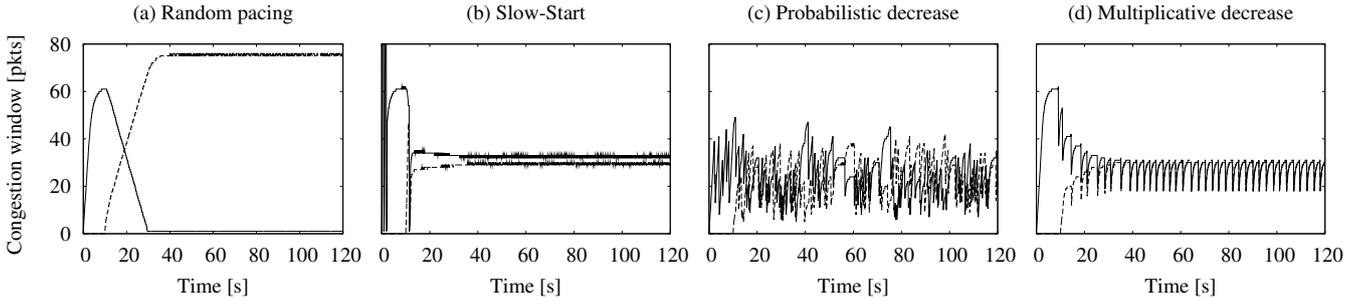

Fig. 2. Effect of the solutions proposed in the two flows scenario.

Afterwards, at time $t = 10$ s a second LEDBAT flow starts. Because of the amount of queuing due to the first flow, the second one measures a base delay equal to $\tau$ resulting in a target twice larger than that of the first flow. So, while the late-comer flow increases its window to reach the target, the first one senses a higher queuing delay and slows down its sending rate. From the queue plot we can see that during this phase the queue never empties: thus, the second flow is never able to correct its wrong base delay estimation and settles after finally reaching its target. At the same time, the first flow enters a starvation phase, which can last for quite a large amount of time until a packet loss occurs or new flows arrive. Moreover, the late-comer issue also extends to multiple flows: as shown in Fig. 1-(c), where a new LEDBAT flow is started every 5 seconds, each new-comer sets a higher target and is always more aggressive than previous flows, which in their turn decrease their throughput to zero.

In the following, we will present a number of mechanisms which try to solve this issue. We address the problem of the wrong estimation of the base delay and investigate the root causes of unfairness that prevent the system from converging to a stable and fair regime.

We use simple metrics to evaluate each solution. On one hand we use *efficiency* ($\eta$), which is the percentage of available bandwidth actually used, that expresses the ability of the protocol to effectively exploit resources. On the other hand, to evaluate the sharing of the resources among flows we use the Jain's index of fairness [17] defined as $F = (\sum_{i=1}^{N} x_i)^2 / (N \sum_{i=1}^{N} x_i^2)$ where $x_i$ is the rate of flow $i$ and $N$ is the number of flows. $F$ is equal to one when resources are equally shared, while is equal to $1/N$ in the worst case where one single flows takes over all resources. In the following we will always refer to *long-term fairness*, considering the share of resource over the whole life of flows. It is often useful to evaluate the *short-term fairness* as well, looking at the protocol behavior at smaller time scales: although we have also evaluated this metric, since the results are similar to the long-term fairness for all experiments, we do not report such results for lack of space.

## IV. ADDRESSING THE MEASUREMENT ERROR

### A. Random pacing

The fairness issue was early identified during the definition of the LEDBAT protocol [5] and was later confirmed by our preliminary simulation studies [12]. Still, according to some of the participants to the draft definition, the randomness present in real networks would somehow prevent the late-comer advantage from showing up: They argue that random delays caused by OSes, routers and background traffic are enough to avoid the queue becoming so stable and the consequent flow synchronization. However, relying on external network conditions to ensure that the protocol actually works is not a good engineering practice. For this reasons, it seemed much more robust to incorporate some randomness in the protocol itself, more specifically to add a random jitter to packet transmission time. In this way the queue is expected to show a much more varying dynamic, thus allowing flows to gather different estimates of the queuing delay, eventually converging to a fair share of the resource.

Since this was officially discussed in the LEDBAT working group, we analyze it as a first solution. We add to our implementation a random pacing module, which randomly spaces the transmission time of packets belonging to a congestion window in the RTT. Each packet is delayed by a random, uniformly chosen interval of time, taking care of avoiding packets reordering.

Fig. 2-(a) shows the same scenario of Fig. 1-(a), but in this case both flows sharing the bottleneck implement random pacing. Unfortunately, only some minor modifications of flow behavior can be observed with respect to the plain LEDBAT situation. First, the increase phase of the second flow is slightly longer, as the perturbations of the queuing delay measurements reflect in the queuing delay and slow down the ramp-up. Second, the late-comer flow attains a slower value of the congestion window, because of its smaller target derived by its different view of the base delay. Nevertheless, random pacing does not constitute a solution, as we assist to the same unfair situation, and we thus disregard it in the following.

### B. Slow start

In our preliminary study [12], we showed how the introduction of a TCP-like slow-start phase at the beginning of a LEDBAT connection has the side-effect of de-correlating flows and allowing them to sense the correct base delay, thus mitigating the unfairness issue. Standard TCP employs such a mechanism in order to converge faster to an optimal utilization of the available bandwidth. However, from our point of view the most important aspect is that a new flow will likely force a loss in the other connections insisting on the same bottleneck. As a consequence all flows will reduce their rate, the queue will be drained empty and all flows, the last one included, will be able to correctly measure the base delay. In Fig. 2-(b) we report an example case, again with two flows which go through a slow start phase at the beginning of their life. As expected the losses due to slow-start of the second flow act as reset: both flows back-off, sample the correct value of base delay and thereafter share the bandwidth equally.

Despite its beneficial effects, the introduction of such an aggressive mechanism in a low priority protocol seems contrary to LEDBAT original design goals. In fact slow-start also disturbs the operation of other protocols sharing the bottleneck, as they will experience losses as well. Though the real number of packet losses can be very limited [12], causing only minor troubles to other services, in the following we try to devise some less intrusive solutions to the fairness issue, which will be anyway inspired by the lesson of slow-start.

## V. INTRODUCING MULTIPLICATIVE DECREASE

From the study of the slow-start solution we can derive a simple intuition: the introduction of a multiplicative decrease in the window dynamics, which causes a sudden drop of sending rate, can relieve the fairness issue. In fact, multiplicative window drops clearly accelerate the buffer drain, thus allowing flows to better estimate the base delay and potentially converge to a stable and fair regime. In other words, we conjecture LEDBAT addictive decrease component to be the principal cause of unfairness.

In this section, after analytically demonstrating the intrinsic instability and unfairness due to the additive decrease component, we explore two ways of explicitly introducing a multiplicative decrease in the LEDBAT protocol: first, we superpose a probabilistic window drop to the LEDBAT linear controller (1) in Sec. V-B; then, we directly replace the additive decrease with a multiplicative one in Sec. V-C.

### A. Impact of additive decrease

We argue that the additive decrease, rather than the measurement errors, is the main cause of unfairness in the LEDBAT protocol: in other words, the late-comer advantage is actually a fundamental drawback of the additive decrease term, meaning that the original design is currently misguided.

Without any loss of generality, let us consider the case of $N$ LEDBAT flows with the same round trip time $R(t)$, sharing the same link of capacity $C$ and finite buffer size $B$. Each flow $i \in \mathcal{N}$, with $\mathcal{N} = \{1, 2, \ldots, N\}$, starts at $t_i \geq 0$, with $t_1 \leq t_2 \leq \cdots \leq t_N$ and with an initial congestion window $W_i$. Given the packet-level congestion window dynamics in (1), we demonstrate the following statement.

**Proposition V.1** *If* $N < \frac{B}{\tau C}$, *and* $d_{max}(t_N) \triangleq \max_{i,j \in \mathcal{N}}[W^i(t_N) - W^j(t_N)] > 0$, *then the system is unfair, i.e.* $\exists t^* \geq t_N$, *such that* $\forall t > t^* \, d_{max}(t) > 0$.

*Proof:* Given (1), a simple fluid representation of the window dynamics of flow $i$, $W_i(t)$, in continuous time, is:

$$\frac{dW_i(t)}{dt} = \frac{1}{R} \frac{\tau - q(t)}{\tau}, \quad (2)$$

where we supposed for simplicity $R(t) \approx R$, which is true for large propagation delay (the proof can be easily extended to the case of variable round trip delays). Since the estimated queuing delay can be different for each flow, depending on its stored base delay, we replace $q(t)$ by $q_i(t)$, i.e., the queue occupancy measured by each sender, and simply observe that $q_i(t)$ varies in the interval $(q(t) - (N-1)\tau, q(t))$. Indeed, the last flow makes the largest error in the estimation of the queuing delay, because it measures as base delay the actual propagation delay increased by $(N-1)\tau$, the sum of the target delay of all preceding flows. It follows that, $\forall i, j \in \mathcal{N}$:

$$W^i(t) - W^j(t) = W^i(t_N) - W^j(t_N) + \int_{t_N}^{t} \frac{q_j(u) - q_i(u)}{R\tau} du \quad (3)$$

where $|q_j(t) - q_i(t)|$ is bounded by $(N-1)\tau$. Hence, if we choose $t^*$ equal to $t_N + \frac{W^{i*}(t_N) - W^{j*}(t_N)}{(N-1)/R}$, with $(i^*, j^*) = \arg\max_{i,j \in \mathcal{N}} W^i(t_N) - W^j(t_N)$, it results:

$$d_{max}(t) \triangleq \max_{i,j \in \mathcal{N}} W^i(t) - W^j(t) > 0, \quad \forall t > t^*.$$

■

It is worth observing that the additive decrease component makes the system not only unfair in general, but also unstable in the Lyapunov sense. This can easily be observed from (3) by looking at the dependence of the regime ($t > t^*$) on the initial condition. Such result has been first shown by Jain in the late 80s [18], in the simpler case when the additive increase/decrease factor is constant and equal for all flows.

### B. Random window dropping

From the above remarks, we learn that a multiplicative decrease must be added for the protocol to work properly: for instance, if LEDBAT flows autonomously slowed down their rate at regular intervals, we could avoid forcing losses in the buffer (i.e., slow-start) altogether. A simple way to induce this behavior is to randomly drop the congestion window: upon reception of an acknowledgment packet, in addition the adjustments specified by (1), we also halve the congestion window with a constant probability $p$. At flow level, this results in a dropping rate proportional to the current transmission rate. The evolution of the congestion window in the simple case of two flows with a drop probability $p = 10^{-4}$ is reported in Fig. 2-(c), showing a fair share.

Now we want to identify an optimal range of values for the drop probability $p$. We preliminary consider the case of two flows arriving at the bottleneck with a gap of $\Delta T = 10\,s$ plus a random jitter uniformly distributed in $[-1, 1]\,\text{ms}$. In

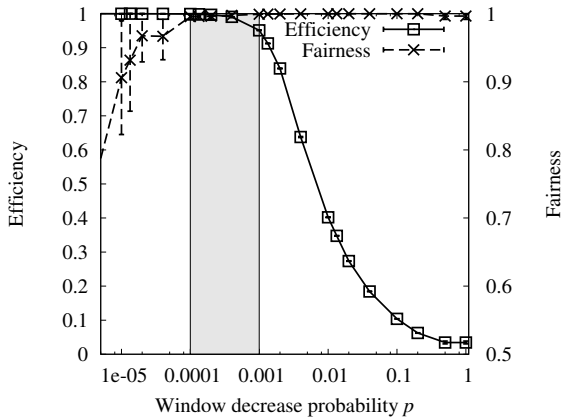

Fig. 3. Efficiency and fairness as a function of $p$

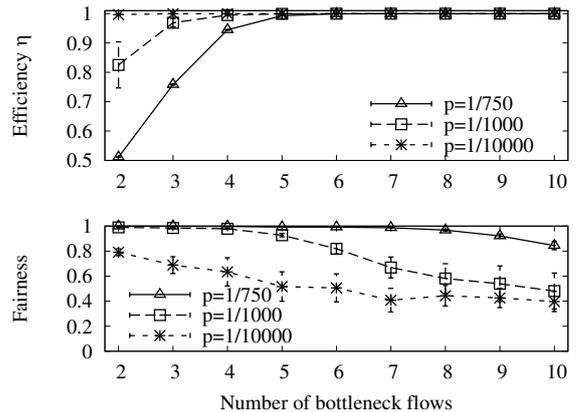

Fig. 4. Peformance of random drop for different number of flows $N$.

Fig. 3, one can observe the resulting resource allocation in terms of efficiency (left axis) and fairness index (right axis) as a function of the chosen drop probability $p$. For each value of $p$ we represent mean and variance (with vertical bars) of the considered metric over 25 simulations, each one lasting 300 s. As expected, for small values of $p$ we obtain a low fairness index (because the drop event is not frequent enough), but an efficient utilization of the bottleneck. On the opposite side, when $p$ becomes high the efficiency is extremely compromised, while the fairness is restored. Despite the natural tradeoff between fairness and efficiency, values of $p$ in the grey-shaded range $[10^{-4}, 10^{-3}]$ seem to allow a fair and efficient share of resources.

Still, the selection of the random probability $p$ strongly depends also on the number of flows sharing the bottleneck: the larger the number of flows, the larger $p$ should be in order to have all flows simultaneously slow down to allow new-comer flows to measure the right base delay. To confirm this intuition we report in Fig. 4 the behavior of $\eta$ and $F$ for three values of $p$ when $N \in [2, 10]$. Mean and variance over 100 simulations of the considered metrics are plotted for the case where each flow starts randomly in $[0, 60]$ s. If the efficiency remains very high, with a good utilization of link starting from $N = 4$ for all $p$ settings, the fairness index shows an improvement over the plain LEDBAT case, but is however far from the optimum. In fact, when multiple flows are involved, one should use a much higher probability to achieve a perfect share of resource, which would in turn impose a more significant cost in term of link efficiency, especially for small values of $N$.

*C. Multiplicative Decrease*

The encouraging results of the previous section and the analysis of Sec.V-A, suggest taking a step further and replacing the LEDBAT additive decrease with a multiplicative one altogether. Therefore, we modify the algorithm so that, whenever an ack packet carries a delay sample exceeding the target $\tau$, the window drops by a factor $\beta < 1$, i.e., $W(t+1) = \beta W(t)$. Notice that the multiplicative decrease is rate-dependent, and thus penalizes flows proportionally to their sending rate (window). Fig. 2-(d) shows the evolution of the congestion window for two competing flows with $\beta = 0.6$. We can observe rate convergence to a stable regime where each flow gets a fair share of the capacity, once both flows have correctly estimated the base delay. Moreover, at steady state flows decrease their window simultaneously: this is a desirable property, since newly arriving flows will have the occasion to correctly measure the propagation delay.

Like in the random drop solution of V-B, a careful choice of the multiplicative factor $\beta$ has to be made. Following the same approach used before, we first study the case of two flows and then consider the general case with a greater number of flows. In Fig. 6 we plot mean and variance over 25 simulations of efficiency and fairness for increasing values of $\beta$ (we actually report the value $1 - \beta$ on the x-axis). As before, flows starts with a gap of $\Delta T = 10\,s$ plus a random jitter. As expected, values of $\beta$ close to 0 (i.e., $1 - \beta$ close to 1) solve the fairness issue but introduces an efficiency loss. On the contrary, a large range of values of $\beta$ (i.e., the gray-shaded part of the $\beta < 0.99$), guarantee both fairness and efficiency.

In the case of $N > 2$ flows, Fig. 5 shows a light dependence of the efficiency on the $\beta$ parameter, with a small efficiency reduction as long as $N$ increases. However this behavior is mainly due to the queuing delay estimation error, whereas the rate-dependent multiplicative decrease is known to asymptotically eliminate such dependence on the number of flows for the TCP case. Intuitively, when multiple flows occupy the buffer, the multiplicative decrease factor should be smaller to obtain a more significant drop in the sending rate and this is exactly what happens since the single flow rate (window) is smaller as $N$ increases.

VI. DISCUSSION AND CONCLUSIONS

In this paper we analyzed four alternative solutions to intra-protocol fairness issues arising in LEDBAT. The first two solutions, based on random pacing and on an additional slow start phase, were respectively inspired by the discussions within the LEDBAT IETF working group [5] and by previous work. Their objective is to de-correlate flow dynamics, so to allow flows to get a correct estimate of the queuing delay. In both cases, unfairness appears to be only partially or ineffectively relieved: the random jitter addition shows no real improvement in terms of fairness, whereas the aggressiveness in window increase of slow-start goes against LEDBAT low-priority goal.

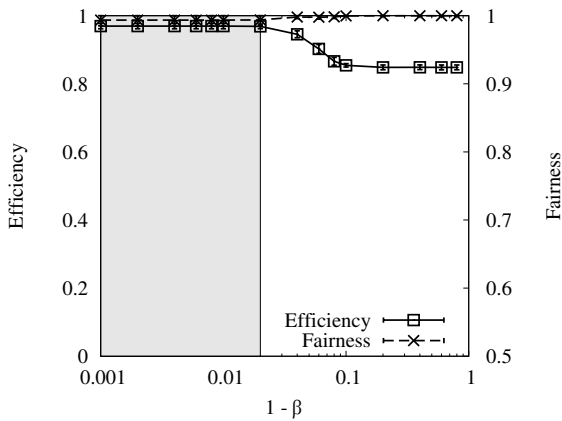

Fig. 5. Efficiency and fairness for different values of $\beta$.

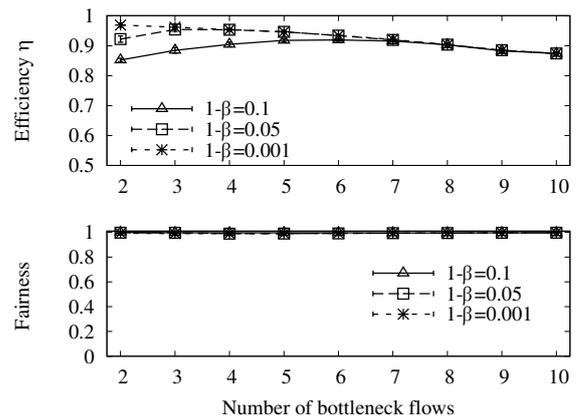

Fig. 6. Performance of multiplicative decrease for different number of flows

While investigating the reasons behind the unfairness, an accurate analysis of window dynamics has highlighted the limits of the additive decrease component. In fact, as already observed in a much simpler scenario by Jain in [18], additive decrease plays an important role in preventing the system from converging to a stable and fair regime. In the LEDBAT case, the error in the estimation of the queuing delay further hinders the convergence to a fair state. Therefore, we devised two possible alternatives to incorporate a multiplicative decrease term in the LEDBAT controller: first, by adding a probabilistic drop to the additive increase/decrease dynamics, then by directly replacing the additive decrease with a multiplicative one altogether. The results are promising as they display a region of the parameters (drop probability $p$ or decrease factor $\beta$) where fairness can be achieved at no or little expense of efficiency.

Although both solutions have their merits, the multiplicative-decrease one may be more appropriate, given: (i) a moderate efficiency loss tolerable in a low-priority context; (ii) the fact that rate-dependent multiplicative decrease attenuates the dependency from the number of competing flows (in fact it is rather due to the measurement error); (iii) the solid theoretical foundation of a purely multiplicative window decrease, in the line of what proved in [18] for constant increase/decrease factors equal for all competing flows. In our future work, we plan to pursue the design of an additive increase/multiplicative decrease controller tailored to LEDBAT goals, carefully addressing the choice of the optimal multiplicative decrease factor $\beta$ and also considering a wider range of scenarios.

ACKNOWLEDGMENTS

This work has been funded by the Celtic project TRANS.